\documentclass[12pt]{article}

\usepackage{sbc-template}
\usepackage{graphicx}
\usepackage{url}
\usepackage[T1]{fontenc}
\usepackage[utf8]{inputenc}
\usepackage[english]{babel}
\usepackage{float}
\usepackage{booktabs}
\usepackage{array}
\usepackage{placeins}
\usepackage{pifont}
\newcommand{\cmark}{\ding{51}}
\newcommand{\xmark}{\ding{55}}

\title{UFU-Do53-EXF and UFU-DoH-EXF: Datasets for DNS Exfiltration Detection\footnote{English version of the paper published at SBSeg 2026 - https://doi.org/10.5753/sbseg.2026.26992}}

\author{Cristiano L. M. Borges\inst{1} \and Rodrigo S. Miani\inst{1}}

\address{Faculty of Computing (FACOM) -- Federal University of Uberlândia (UFU)
\email{\{cristiano.borges,miani\}@ufu.br}
}

\begin{document}

\maketitle

\begin{abstract}
Data exfiltration through the \textit{Domain Name System} (DNS) remains a relevant cybersecurity challenge, as DNS is an essential network service that can be misused as a covert channel for unauthorized data transfer. This paper introduces two complementary datasets, \textit{UFU-Do53-EXF} and \textit{UFU-DoH-EXF}, designed to support the study of data exfiltration in conventional DNS over port 53 and DNS over HTTPS. The proposed methodology combines controlled generation of benign and malicious traffic, PCAP packet captures, a reproducible \textit{PCAP-to-Flow} processing pipeline, and protocol-specific feature extraction. After normalization, \textit{UFU-Do53-EXF} comprises 5,351,064 samples, whereas \textit{UFU-DoH-EXF} comprises 1,048,575 TCP/443 flows. Together, the datasets support the analysis of exfiltration behavior in both clear-text and encrypted DNS traffic. Supervised baseline evaluations using widely adopted machine learning classifiers indicate that the proposed representations are consistent and capture discriminative patterns for distinguishing benign traffic from exfiltration activity.
\end{abstract}

\section{Introduction}

Data exfiltration refers to the unauthorized transfer of information from a compromised environment to attacker-controlled infrastructure. In knowledge bases widely adopted in the field, this stage is recognized as a recurring mechanism for carrying out exfiltration in intrusion campaigns, including through command-and-control channels or the use of alternative communication protocols \cite{mitre_attck_t1041_2026,mitre_attck_t1048_2026,mitre_attck_t1048003_2026}. Among the protocols susceptible to this abuse, DNS stands out because it is indispensable to the operation of network services and, in many environments, less restricted than other egress channels \cite{steadman_dnsxd_2018}.

In traditional DNS, exfiltration can occur by encoding data in subdomains, by tunneling, or by using the protocol as a covert channel to transmit fragments of information \cite{steadman_dnsxd_2018,ahmed_real-time_2019,abualghanam_real-time_2023}. In \textit{DNS over HTTPS} (DoH), detection becomes more difficult because DNS messages are encapsulated in HTTPS/TLS, usually over HTTP/2, which reduces direct content inspection and shifts the analysis toward flow statistics, packet direction, observable sizes, and temporal behavior \cite{hoffman_dns_2018,rescorla_transport_2018,zhan_detecting_2022}.

Recent literature has shown that detecting DNS exfiltration with machine learning is feasible \cite{elaoumari_evasion-resilient_2025}, but it remains conditioned on the quality of the datasets used, the traceability of the labeling process, and the alignment between the sampling unit and the observed phenomenon \cite{sabir_machine_2021,garcia_towards_2023}. In this context, this work presents two datasets: \textit{UFU-Do53-EXF}, focused on conventional DNS, and \textit{UFU-DoH-EXF}, focused on DNS over HTTPS. Both were generated under the same experimental design, with controlled generation of benign and malicious traffic, preservation of artifacts as PCAP files and network flows, explicit feature definitions, and methodological documentation.

The main contributions of this work are:
\begin{itemize}
    \item the proposal of two datasets specifically oriented toward DNS exfiltration detection, covering the Do53 and DoH scenarios;
    \item the definition of a common experimental framework, with controlled generation of benign and malicious traffic, preservation of PCAPs, reproducible conversion to network flows, and explicit labeling;
    \item the characterization of features specific to clear-text DNS and to DNS over HTTPS, respecting the differences in observability between the protocols;
    \item the presentation of baseline supervised evaluations for the \textit{UFU-Do53-EXF} and \textit{UFU-DoH-EXF} datasets, using classifiers widely employed in detection problems.
\end{itemize}

\section{Data Exfiltration via DNS}

DNS exfiltration exploits the constant presence of this protocol in network communications and its operational importance for name resolution. Because it takes part in a large share of the interactions among clients, applications, and services, DNS offers attackers a means of transporting data without necessarily resorting to channels that are more closely scrutinized by traditional inspection mechanisms. As a result, the protocol has come to occupy a prominent position in the attack surface, both in command-and-control operations and in tunneling and data exfiltration scenarios \cite{mitre_attck_t1041_2026,mitre_attck_t1048_2026,mitre_attck_t1048003_2026}.

In its most common form, DNS exfiltration consists of fragmenting the content of interest, applying a suitable encoding, and inserting these fragments into queries directed to a domain under the adversary's control. The authoritative infrastructure associated with that domain can then reassemble the received data. This mechanism benefits from the flexibility of DNS, especially in the use of subdomain labels, and can be adapted to reduce suspicion by adjusting the frequency, size, and distribution of queries \cite{steadman_dnsxd_2018,gunes_gursoy_dns_2024}.

In the Do53 context, visibility into the content of the DNS message makes it possible to exploit features associated with the structure of the queried name and with resolution behavior. Query length, domain name depth, label entropy, digit ratio, record types, and per-domain aggregate statistics are useful signals for distinguishing benign traffic from artificial tunneling and exfiltration patterns \cite{ahmed_monitoring_2020,tang_practical_2020,palau_detecting_2019,nadler_detection_2019}. In benign traffic, it is more common to observe regular names, recurring domains associated with well-known applications, and coherent combinations of query types. In malicious traffic, by contrast, encoded strings, lexical irregularity, and patterns produced for data transport become more frequent \cite{abualghanam_real-time_2023,ahmed_real-time_2019}.

From a defensive perspective, approaches based exclusively on blocklists have important limitations. Static blocklists can be useful as a complementary mechanism, but they tend to respond inadequately to new domains, rotating infrastructure, and malicious queries that attempt to mimic legitimate behavior. For this reason, effective defense requires in-depth query inspection, contextual analysis, and detection of anomalous patterns related to the length, frequency, volume, entropy, and lexical structure of the queried names \cite{ahmed_real-time_2019,ahmed_monitoring_2020,steadman_dnsxd_2018}. This perspective is consistent with the understanding that successful attacks evolve faster than purely reactive blocking mechanisms can keep up with.

In DoH, the nature of the problem changes substantially. Since DNS content is encapsulated in HTTPS/TLS, direct inspection of queries is no longer available in clear text to traditional monitoring mechanisms. Consequently, detection must rely on side effects of communication behavior, such as flow duration, packet counts, size statistics in each direction, inter-packet intervals, and temporal information associated with TLS connection establishment and data transmission \cite{hoffman_dns_2018,rescorla_transport_2018,zhan_detecting_2022,moure-garrido_real_2023,salat_dns_2023}. In practical terms, the spread of encrypted DNS increases monitoring difficulty and imposes greater reliance on flow statistics, temporal patterns, and complementary controls for detection and security policy enforcement \cite{zhan_detecting_2022,moure-garrido_real_2023}.

This distinction between Do53 and DoH is methodologically relevant because it directly affects how the detection problem is formulated. In the former, modeling can exploit features derived from the DNS query itself. In the latter, the analysis must shift to indirect signals associated with encrypted transport. In both scenarios, however, DNS exfiltration remains a technically viable form of protocol abuse, which justifies building specific and methodologically consistent datasets to support detection studies \cite{sabir_machine_2021,steadman_dnsxd_2018,zhan_detecting_2022}.

\section{Related Work}

\subsection{DNS Exfiltration Detection Methods}


DNS exfiltration detection depends on visibility into the protocol and on the features analyzed. In conventional DNS over port 53 (Do53), where query content remains observable, approaches based on lexical domain features, protocol metadata, and aggregate behavioral statistics are common. Previous works range from real-time detection in enterprise networks to the use of supervised and deep learning to identify covert communications and threats associated with DNS abuse \cite{ahmed_real-time_2019,tang_practical_2020,palau_detecting_2019}. Other studies highlight challenges such as low-throughput exfiltration, stealthy tunneling, and the limitations of volumetric metrics, reinforcing the importance of structural features, entropy, and contextual information about the queried domains \cite{nadler_detection_2019,steadman_dnsxd_2018,abualghanam_real-time_2023}.

Overall, the methods proposed for Do53 converge on the idea that DNS abuse tends to produce discriminative signals observable in the query content itself or in associated resolution patterns. Name length, hierarchical depth, character distribution, presence of encoded strings, digit ratio, and record types are among the most frequently exploited signals for distinguishing benign traffic from artificial tunneling and exfiltration behavior \cite{ahmed_monitoring_2020,tang_practical_2020,nadler_detection_2019,palau_detecting_2019}. This line of work shows that supervised detection can achieve good results when the data representation adequately preserves the relevant properties of DNS communication.

In the DoH context, the problem takes on distinct characteristics, since DNS messages are encapsulated in HTTPS/TLS, which reduces the possibility of direct semantic inspection of queries. In this scenario, the literature shifts modeling toward indirect features, such as flow duration, packet count, per-direction size statistics, temporal intervals, and information associated with TLS connection establishment and data transmission. Zhan et al. demonstrate that exfiltration detection in DoH remains feasible even without access to clear-text DNS content, provided that suitable statistical features are used \cite{zhan_detecting_2022}. Similarly, Moure-Garrido et al. and Salat et al. show that malicious DoH traffic tends to introduce measurable changes in temporal and directional patterns, although encryption imposes additional limitations on the interpretability of the results \cite{moure-garrido_real_2023,salat_dns_2023}.

More broadly, reviews and comparative studies indicate that the performance of detection methods depends not only on the algorithm used, but also on labeling quality, on the alignment between the sampling unit and the observed phenomenon, and on the methodological consistency of data collection \cite{sabir_machine_2021,garcia_towards_2023,luz_unsupervised_2023,ozery_information-based_2024}. Thus, although there is consistent evidence that DNS exfiltration detection is feasible in both Do53 and DoH, the practical usefulness of these methods remains strongly conditioned on the availability of well-documented, representative datasets compatible with the protocol under analysis \cite{sabir_machine_2021,garcia_towards_2023}.

\subsection{Datasets for DNS Exfiltration Detection}

Beyond the evolution of detection methods, the literature also highlights the importance of the datasets used in experimental evaluation. In the case of DNS exfiltration, this issue is particularly relevant because different public datasets adopt distinct objectives, non-equivalent sampling units, and varying levels of methodological documentation. As a consequence, comparisons across studies do not always reflect only differences between models, but also structural differences between the data themselves \cite{garcia_towards_2023,sabir_machine_2021}.

Among the datasets frequently cited in the literature, the \textit{DNS Exfiltration Dataset} gathers a large volume of samples focused on exfiltration in traditional DNS, but does not provide, in a structured manner, all the artifacts required to fully trace the data generation and transformation process. \textit{CIC-Bell-DNS-EXF-2021} is an important reference for Do53, combining capture artifacts and network flow representations, although it was not designed to establish a direct methodological parallel with the encrypted DNS scenario. \textit{CIRA-CIC-DoHBrw-2020}, in turn, stands out as a relevant dataset for DoH studies, providing captures, flows, and scripts, but with a focus distinct from the specific problem of exfiltration under the same experimental design adopted for conventional DNS.

Table~\ref{tab:datasets_comparison} summarizes this comparison between related datasets and the datasets proposed in this work. The goal is not only to contrast sample volume, but to highlight criteria that directly affect the scientific usefulness of a dataset, such as protocol coverage, availability of PCAP artifacts, network flow representation, and availability of scripts that support reproducibility.

\begin{table}[htbp]
\centering
\caption{Comparison between related datasets and the proposed datasets.}
\label{tab:datasets_comparison}
\scriptsize
\setlength{\tabcolsep}{3pt}
\renewcommand{\arraystretch}{1.15}
\begin{tabular}{>{\raggedright\arraybackslash}p{2.8cm}ccccccc}
\toprule
\textbf{Dataset} & \textbf{Do53} & \textbf{DoH} & \textbf{PCAP} & \textbf{Flow} & \textbf{Scripts} & \textbf{Comparability} & \textbf{Samples} \\
\midrule
DNS Exfiltration Dataset & \cmark & \xmark & \xmark & \cmark & \xmark & \xmark & 1,583,163 \\
CIC-Bell-DNS-EXF-2021 & \cmark & \xmark & \cmark & \cmark & \xmark & \xmark & 1,019,318 \\
CIRA-CIC-DoHBrw-2020 & \xmark & \cmark & \cmark & \cmark & \cmark & \xmark & 269,643 \\
\textbf{UFU-Do53-EXF} & \cmark & \xmark & \cmark & \cmark & \cmark & \cmark & 5,351,064 \\
\textbf{UFU-DoH-EXF} & \xmark & \cmark & \cmark & \cmark & \cmark & \cmark & 1,048,575 \\
\bottomrule
\end{tabular}
\end{table}

It can be observed that existing datasets offer important contributions, but also present recurring limitations when considered from the perspective of methodological comparison between Do53 and DoH. In many cases, the datasets cover only one of the scenarios, use different sampling units, do not simultaneously preserve captures and structured representations, or do not provide sufficient documentation to accurately reconstruct the experimental process. These limitations affect not only reproducibility, but also comparability across studies that seek to understand the impact of the change in observability between traditional DNS and encrypted DNS. In this context, the motivation for creating the \textit{UFU-Do53-EXF} and \textit{UFU-DoH-EXF} datasets stems from the need to narrow this methodological gap. By establishing a common experimental design, with controlled generation of benign and malicious traffic, preservation of PCAP and network flow artifacts, consistent labeling, and features aligned with the characteristics of each protocol, we seek to provide a more consistent basis for the comparative study of DNS exfiltration in Do53 and DoH scenarios \cite{garcia_towards_2023,sabir_machine_2021,srivastava_xai_2022}.

\section{Description of the Proposed Datasets}

This section describes the construction process of the \textit{UFU-Do53-EXF} and \textit{UFU-DoH-EXF} datasets. The methodology was designed to maintain coherence between the two protocols analyzed, preserving comparability between benign and malicious scenarios, traceability of traffic generation, and consistency in transforming captures into representations suitable for modeling. In both cases, the construction of the datasets started from the same experimental design, with prior definition of the traffic scenarios, a reproducible laboratory topology, capture in PCAP format, and subsequent conversion into structured flow-level representations. To enable the reproducibility of the experiments, all artifacts, scripts, and methodological procedures used in this study have been made publicly available in the project's GitHub repository: \url{https://github.com/CborgesUFU/DNS-360-Dataset}.

\subsection{Overview of the UFU-Do53-EXF and UFU-DoH-EXF Datasets}

The proposed datasets were organized in a complementary manner. \textit{UFU-Do53-EXF} was designed to represent the exfiltration problem in conventional DNS, preserving features directly observable in the queries and responses exchanged over port 53. \textit{UFU-DoH-EXF}, in turn, was built to represent the same problem in a context of lower semantic observability, in which DNS messages are encapsulated in HTTPS/TLS and the analysis depends predominantly on flow statistics and temporal information of the TLS traffic.

Each dataset incorporates two main scenarios. The first corresponds to benign traffic, generated in a controlled manner to represent legitimate resolutions compatible with the protocol analyzed. The second corresponds to malicious exfiltration traffic, produced by specific tools and routines for covert data transport. This organization into benign and malicious scenarios was maintained for both Do53 and DoH, allowing distinct protocols to be compared under the same methodological framework. In \textit{UFU-Do53-EXF}, the benign scenario consists of traditional DNS queries directed to legitimate resolvers, with diversity of domains and record types. The malicious scenario, in turn, involves data exfiltration through manipulated DNS queries, in which fragments of information are encapsulated in domain names and forwarded to attacker-controlled infrastructure. In \textit{UFU-DoH-EXF}, the benign scenario consists of DoH queries performed through HTTPS/TLS requests to public resolvers, whereas the malicious scenario consists of a custom client capable of exfiltrating data through requests compatible with the \textit{DNS over HTTPS} paradigm.

\subsection{Experimental Parameters}

The construction of the benign scenarios started from a controlled selection of domains, with the aim of reducing selection bias and favoring reproducibility. To this end, the benign queries were based on domains extracted from the \textit{Tranco Top 1 Million List}, designed specifically for research use and hardened against manipulation of the ranking \cite{le_pochat_tranco_2019}. There is no guarantee that the included domains are benign, since their selection is based on popularity rather than on security or reputation criteria. Nevertheless, this choice helps avoid arbitrary or poorly documented lists and improves the quality of the benign baseline in both Do53 and DoH.

In addition to domain selection, generating the exfiltration scenarios required defining a heterogeneous set of files capable of representing artifacts commonly found on workstations. For this purpose, a set of audio, video, image, text, binary executable, and compressed files was used. These files were organized into two profiles, called \textit{light} and \textit{heavy}, in order to induce controlled variations in the volume of exfiltrated data. For experimental organization purposes, files up to approximately 1 MB in size were considered \textit{light}, and files larger than this value were considered \textit{heavy}. This strategy made it possible to produce malicious traffic with different intensities without compromising consistency between the experiments conducted in Do53 and DoH.

Adopting the same logical inputs for both protocols was important to reduce external interference in the comparison. Rather than simultaneously varying file type, protocol, and topology, we chose to keep the artifacts used in the exfiltration fixed and to vary only the transport and observation mechanisms corresponding to each protocol. This made it more feasible to compare the two contexts without introducing unnecessary methodological differences.

\subsection{Laboratory Topology}

The captures were generated in a laboratory topology implemented in a cloud environment, on Amazon Web Services (AWS). The laboratory design was guided by three main objectives: isolating the attacker and victim roles in distinct segments, ensuring reproducibility of addressing and connectivity rules, and enabling collection at complementary points of the communication. The infrastructure was organized as a \textit{Virtual Private Cloud} (VPC) logically segmented into two private networks. One network corresponds to the attacker's environment and the other to the victim's environment. This separation was adopted to keep the assignment of experimental roles clear and to facilitate repeating the scenarios under controlled conditions.

\begin{figure}[htbp]
    \centering
    \includegraphics[width=\linewidth]{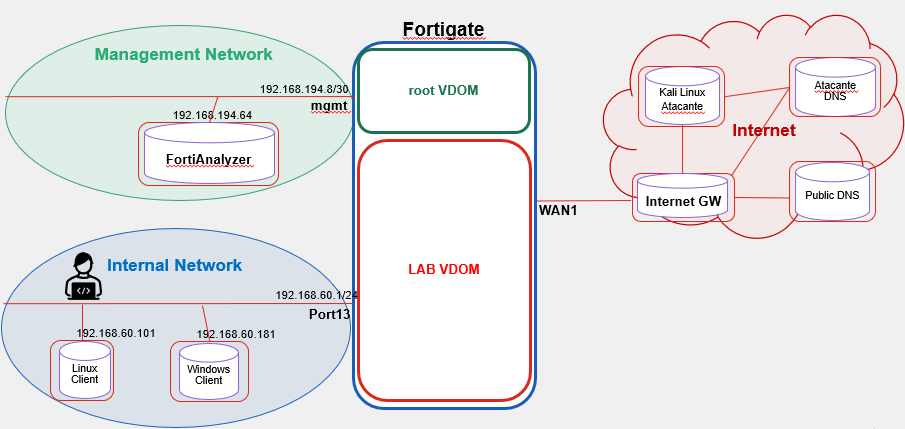}
    \caption{Laboratory topology used to generate the Do53 and DoH scenarios.}
    \label{fig:topologia_lab}
\end{figure}

\FloatBarrier

As illustrated in Figure~\ref{fig:topologia_lab}, among the laboratory components, a FortiGate with an \textit{Evaluation License} was used as the routing and traffic collection element, supporting the observation of communications and, when necessary, capture through complementary mechanisms. FortiAnalyzer was used for log consolidation and to support event correlation. The environment also included a Windows server for auxiliary functions, an attacker machine based on Kali Linux, and a victim machine based on Ubuntu. The attacker machine ran the exfiltration and command-and-control tools, including \texttt{dnscat2} for the Do53 scenario and the client or routines for the DoH scenario. This topology, although deliberately reduced compared to real enterprise environments, was sufficient to reproduce benign and malicious flows in Do53 and DoH, record the communications in PCAP format, and support the subsequent stage of transformation into flow-level representations.

\subsection{Traffic Scenarios and Capture Generation}

Capture generation was organized into four experimental scenarios. Two of them correspond to traditional DNS and two to \textit{DNS over HTTPS}, each pair comprising one benign scenario and one exfiltration scenario. This organization makes it possible to evaluate traffic behavior in two distinct observability contexts while preserving a common experimental logic.

In the benign Do53 scenario, the victim machine performs conventional DNS queries over 53/UDP to legitimate resolvers. The goal of this scenario is to establish a controlled baseline for name resolution behavior. To avoid an overly restricted benign scenario, both the volume and the timing of the queries were controlled by dedicated scripts, and the diversity of the queried destinations was guided by the Tranco list \cite{le_pochat_tranco_2019}. In addition, different record types may be used throughout generation, increasing the variability of the observable protocol fields. In the malicious Do53 scenario, the victim machine establishes a command-and-control channel over DNS using the \texttt{dnscat2} tool. In this configuration, data are fragmented and encapsulated in domain names, which are then transmitted in DNS queries directed to attacker-controlled infrastructure. This tool was well suited to the methodological objective of the work, as it allows repeatability, experimental control, and the generation of traffic consistent with the literature on DNS tunneling and exfiltration \cite{bowes_dnscat2_2022,steadman_dnsxd_2018}.

In the benign DoH scenario, the victim performs queries to public \textit{DNS over HTTPS} resolvers through HTTPS/TLS requests on port 443. In this case, traffic observation no longer relies on clear-text DNS content, but on side effects of the encrypted transport, such as flow duration, packet counts, sizes in each direction, flow statistics, and temporal information associated with TLS connection establishment and data transmission. To avoid a biased baseline, diversity of domains and resolvers was pursued while preserving the controlled nature of the collection. In the malicious DoH scenario, a custom client exfiltrates the files through HTTPS requests compatible with the \textit{DNS over HTTPS} paradigm. The content is fragmented and distributed across multiple requests, so that the effect of the exfiltration manifests mainly in the bidirectional flow statistics and in the temporal information associated with the TLS traffic, rather than in the content of the DNS messages themselves. This scenario is particularly relevant because it represents a condition of lower semantic observability, closer to the challenges posed by encrypted DNS in operational environments.

\subsection{PCAP-to-Flow Conversion Pipeline}

The conversion of the network captures into the final datasets was carried out through a reproducible \textit{PCAP-to-Flow} processing pipeline implemented in Python. The purpose of this stage was to transform PCAP/PCAPNG files into flows suitable for the machine learning experiments. Although the \textit{UFU-Do53-EXF} and \textit{UFU-DoH-EXF} datasets were built using protocol-specific routines, both followed the same methodological logic. The implementation used the \texttt{dpkt} library to read and parse the PCAP/PCAPNG files. The \texttt{NumPy} and \texttt{pandas} libraries were used, respectively, to compute the numerical statistics and to organize the extracted samples in tabular form. This choice kept the pipeline independent of proprietary tools and favored the reproducibility of the process of transforming the captures into structured data.

In the Do53 scenario, the pipeline processes conventional DNS packets associated with port 53, covering traffic over UDP and TCP when applicable. For each valid DNS message, features directly related to the queried name, the query metadata, and the observed responses are extracted. In the DoH scenario, the pipeline processes TCP communications associated with port 443. Since DNS content is encapsulated in HTTPS/TLS, the extraction does not depend on direct inspection of DNS queries. Instead, packets are organized into bidirectional flows, considering the client$\rightarrow$server and server$\rightarrow$client directions. For each sampling unit, packet size statistics, temporal intervals, number of packets per direction, communication duration, and total byte volume are computed.

The samples were labeled according to the experimental scenario from which each capture originated. Files associated with the benign scenarios received the \textit{Benign} label, whereas files generated in the exfiltration scenarios received the \textit{Exfiltration} label. This strategy was adopted because the traffic was produced in a controlled environment, with explicit separation between benign and malicious scenarios. Thus, the label of each sample derives from the experimental procedure that produced the PCAP/PCAPNG file, preserving traceability among capture, scenario, and tabular instance.

Therefore, the \textit{PCAP-to-Flow} pipeline is a central stage of the proposed methodology. It ensures that the \textit{UFU-Do53-EXF} and \textit{UFU-DoH-EXF} datasets are derived from raw network captures through an explicit, traceable, and reproducible process. At the same time, it preserves the specificity of each protocol: in Do53, it exploits features directly observable in DNS messages, and in DoH, it uses flow statistics compatible with the encrypted traffic scenario.

\subsection{Do53 Features}

The 29 features of \textit{UFU-Do53-EXF} were defined based on the feature set proposed by Mahdavifar et al.~\cite{mahdavifar_lightweight_2021}, which presents a hybrid, lightweight approach for detecting DNS data exfiltration by combining \textit{stateless} and \textit{stateful} features. In the present work, this set was adapted to the adopted experimental design and to the structure of the generated artifacts, covering lexical properties of the queried name, DNS metadata, and per-domain aggregate statistics. This choice is consistent with the literature that associates DNS abuse with lexical irregularity, increased entropy, the presence of artificial encoding in subdomain labels, and unusual resolution patterns \cite{ahmed_monitoring_2020,tang_practical_2020,palau_detecting_2019,nadler_detection_2019,mahdavifar_lightweight_2021}. Table~\ref{tab:features_do53_summary} presents the feature set adopted for \textit{UFU-Do53-EXF}.

\begin{table}[htbp]
\centering
\tiny
\renewcommand{\arraystretch}{1.15}
\setlength{\tabcolsep}{4pt}
\begin{tabular}{|l|p{9.5cm}|}
\hline
\textbf{Feature} & \textbf{Description} \\
\hline
\texttt{query\_length} & Length of the fully qualified domain name (\textit{FQDN}). \\
\texttt{subdomain\_length} & Length of the subdomain, corresponding to the first label of the \textit{FQDN}. \\
\texttt{num\_labels} & Number of labels in the \textit{FQDN}. \\
\texttt{max\_label\_length} & Maximum length among the labels of the \textit{FQDN}. \\
\texttt{avg\_label\_length} & Average length of the labels of the \textit{FQDN}. \\
\texttt{entropy} & Entropy of the \textit{FQDN}, used as a measure of character randomness. \\
\texttt{ratio\_digits} & Proportion of digits in the queried name, excluding dots. \\
\texttt{ratio\_uppercase} & Proportion of uppercase characters in the queried name, excluding dots. \\
\texttt{ratio\_lowercase} & Proportion of lowercase characters in the queried name, excluding dots. \\
\texttt{ratio\_special\_chars} & Proportion of non-alphanumeric characters, excluding dots and hyphens. \\
\texttt{rr\_type} & Type of the DNS record queried or returned. \\
\texttt{query\_class} & Class of the DNS query. \\
\texttt{ttl} & \textit{Time To Live} value observed in the returned records, when applicable. \\
\texttt{contains\_base64\_pattern} & Indicator of the presence of a pattern compatible with Base32/Base64 encoding in the domain. \\
\texttt{contains\_hex\_pattern} & Indicator of the presence of long hexadecimal sequences in the domain. \\
\texttt{contains\_dictionary\_word} & Indicator of the occurrence of common words. \\
\texttt{ratio\_vowels} & Proportion of vowels in the queried name. \\
\texttt{ratio\_consonants} & Proportion of consonants in the queried name. \\
\texttt{num\_distinct\_characters} & Number of distinct characters in the queried name. \\
\texttt{mean\_query\_length\_per\_domain} & Mean length of the queries grouped by base domain. \\
\texttt{std\_query\_length\_per\_domain} & Standard deviation of the length of the queries grouped by base domain. \\
\texttt{mean\_entropy\_per\_domain} & Mean entropy of the queries grouped by base domain. \\
\texttt{num\_unique\_ips} & Number of distinct IP addresses observed per base domain. \\
\texttt{num\_unique\_asns} & Number of distinct autonomous systems observed per base domain. \\
\texttt{num\_unique\_ttls} & Number of distinct TTL values observed per base domain. \\
\texttt{mean\_ttl} & Mean TTL value per base domain. \\
\texttt{var\_ttl} & TTL variance per base domain. \\
\texttt{num\_queries\_per\_domain} & Total number of queries observed per base domain. \\
\texttt{ratio\_failed\_queries} & Proportion of failed queries per base domain. \\
\hline
\end{tabular}
\caption{Set of 29 features defined for \textit{UFU-Do53-EXF}.}
\label{tab:features_do53_summary}
\end{table}

\subsection{DoH Features}


The 28 features of \textit{UFU-DoH-EXF} were defined based on an adaptation of the approach of Zhan et al.~\cite{zhan_detecting_2022}, aimed at detecting exfiltration via DNS over HTTPS. Since, in DoH, the semantic content of DNS queries and responses is not directly available for inspection, the authors exploit observable characteristics of the communication, such as TLS record sizes, temporal intervals, and packet direction. In this work, this logic was adapted to \textit{UFU-DoH-EXF}, prioritizing flow features related to packet direction, observable sizes, duration, volume, and temporal statistics.

With the standardization of TLS Encrypted Client Hello by RFC 9849 in 2026, part of the information traditionally extracted from the clear-text \textit{ClientHello} tends to become unavailable to passive inspection mechanisms. In particular, the real SNI, the real ALPN, and the JA3 fingerprint associated with the effective \textit{ClientHello} may become unavailable, since they become part of the encrypted \textit{ClientHelloInner}. In these scenarios, network sensors will observe only the \textit{ClientHelloOuter}, which may contain public, generic, or innocuous values, reducing the reliability of these features for classification. Therefore, approaches for detecting exfiltration via DoH should prioritize features that are independent of the \textit{ClientHello} content, such as flow statistics, direction, size, duration, volume, and temporal behavior \cite{rescorla_tls_2026}.

This methodological choice is consistent with the premise that, in DoH scenarios, detection should exploit indirect signals of communication behavior rather than depend on direct inspection of the queried domain names. Accordingly, the features of \textit{UFU-DoH-EXF} were structured to capture patterns associated with the directional distribution of packets, size variations, temporal regularity, flow duration, and observed volume, elements that can differentiate benign traffic from flows used for data exfiltration. Table~\ref{tab:features_doh_summary} presents the feature set adopted for \textit{UFU-DoH-EXF}, and Table~\ref{tab:dataset_distribution} presents the quantitative summary of the two datasets built.

\begin{table}[htbp]
\centering
\tiny
\renewcommand{\arraystretch}{1.15}
\setlength{\tabcolsep}{4pt}
\begin{tabular}{|l|p{9.5cm}|}
\hline
\textbf{Feature} & \textbf{Description} \\
\hline
\texttt{src2dst\_min\_length} & Minimum packet size in the client$\rightarrow$server direction. \\
\texttt{src2dst\_max\_length} & Maximum packet size in the client$\rightarrow$server direction. \\
\texttt{src2dst\_mean\_length} & Mean packet size in the client$\rightarrow$server direction. \\
\texttt{src2dst\_stddev\_length} & Standard deviation of packet size in the client$\rightarrow$server direction. \\
\texttt{dst2src\_min\_length} & Minimum packet size in the server$\rightarrow$client direction. \\
\texttt{dst2src\_max\_length} & Maximum packet size in the server$\rightarrow$client direction. \\
\texttt{dst2src\_mean\_length} & Mean packet size in the server$\rightarrow$client direction. \\
\texttt{dst2src\_stddev\_length} & Standard deviation of packet size in the server$\rightarrow$client direction. \\
\texttt{bidirectional\_max\_length} & Largest packet size considering both directions of the flow. \\
\texttt{bidirectional\_mean\_length} & Mean packet size considering both directions of the flow. \\
\texttt{bidirectional\_stddev\_length} & Standard deviation of packet size in the bidirectional flow. \\
\texttt{src2dst\_min\_time} & Smallest inter-packet interval in the client$\rightarrow$server direction. \\
\texttt{src2dst\_max\_time} & Largest inter-packet interval in the client$\rightarrow$server direction. \\
\texttt{src2dst\_mean\_time} & Mean inter-packet interval in the client$\rightarrow$server direction. \\
\texttt{src2dst\_stddev\_time} & Standard deviation of the inter-packet interval in the client$\rightarrow$server direction. \\
\texttt{dst2src\_min\_time} & Smallest inter-packet interval in the server$\rightarrow$client direction. \\
\texttt{dst2src\_max\_time} & Largest inter-packet interval in the server$\rightarrow$client direction. \\
\texttt{dst2src\_mean\_time} & Mean inter-packet interval in the server$\rightarrow$client direction. \\
\texttt{dst2src\_stddev\_time} & Standard deviation of the inter-packet interval in the server$\rightarrow$client direction. \\
\texttt{bidirectional\_min\_time} & Smallest temporal interval observed in the bidirectional flow. \\
\texttt{bidirectional\_max\_time} & Largest temporal interval observed in the bidirectional flow. \\
\texttt{bidirectional\_mean\_time} & Mean temporal interval considering the bidirectional flow. \\
\texttt{bidirectional\_stddev\_time} & Standard deviation of the temporal intervals in the bidirectional flow. \\
\texttt{num\_src2dst} & Number of packets in the client$\rightarrow$server direction. \\
\texttt{num\_dst2src} & Number of packets in the server$\rightarrow$client direction. \\
\texttt{flow\_duration} & Total duration of the TLS flow. \\
\texttt{byte\_count} & Total bytes observed in the bidirectional flow. \\
\texttt{handshake\_time} & Approximate TLS handshake time. \\
\hline
\end{tabular}
\caption{Set of 28 features defined for \textit{UFU-DoH-EXF}.}
\label{tab:features_doh_summary}
\end{table}

Compared with the feature set originally explored by Zhan et al.~\cite{zhan_detecting_2022}, this work preserves the emphasis on size, direction, and temporal interval statistics, but incorporates additional features related to packet count, duration, volume, and approximate communication establishment time. Specifically, the features \texttt{num\_src2dst}, \texttt{num\_dst2src}, \texttt{flow\_duration}, \texttt{byte\_count}, and \texttt{handshake\_time} were added. This extension is justified by the experimental design adopted in this work, in which the DoH samples are extracted from time windows within the TCP/443 communications. This choice reduces the risk that long connections aggregate heterogeneous behaviors into a single instance and makes it possible to capture local variations associated with the intensity, regularity, and asymmetry of exfiltration traffic.

\begin{table}[H]
\centering
\caption{Distribution of samples in the proposed datasets.}
\label{tab:dataset_distribution}
\footnotesize
\begin{tabular}{lrrrrr}
\toprule
\textbf{Dataset} & \textbf{Features} & \textbf{Samples} & \textbf{Benign} & \textbf{Exfiltration} & \textbf{\% Exfiltration} \\
\midrule
\textit{UFU-Do53-EXF} & 29 & 5,351,064 & 4,812,119 & 538,945 & 10.0706\% \\
\textit{UFU-DoH-EXF}  & 28 & 1,048,575 & 1,016,438 & 32,137 & 3.0648\% \\
\bottomrule
\end{tabular}
\end{table}

\section{Case Studies}

The case studies aim to provide a baseline supervised evaluation for the proposed datasets. In both scenarios, a stratified 80/20 split was adopted, preserving the proportion between the classes. The metrics were chosen considering the class imbalance: in addition to accuracy, balanced accuracy, macro-averaged metrics, and the absolute numbers of false positives and false negatives are reported. This choice prevents performance from being interpreted solely on the basis of the majority class. Furthermore, it is important to note that the two evaluations presented here are intra-dataset experiments based on random stratified splits. Therefore, the results should be interpreted as an initial assessment of the consistency and discriminative capacity of the extracted features, and not as definitive evidence of generalization to other environments.

\subsection{\textit{UFU-Do53-EXF}}


The first case study evaluates \textit{UFU-Do53-EXF} in a supervised binary classification task between \textit{Benign} and \textit{Exfiltration} traffic. The \textit{Random Forest}, \textit{XGBoost}, and \textit{Support Vector Machine} (SVM) models were used, as they are widely employed in machine-learning-based detection and in studies on DNS exfiltration \cite{mahdavifar_lightweight_2021}.

The experiments used a stratified split of 80\% for training and 20\% for testing, preserving the proportion between the classes. The test set contains 1,070,214 instances, of which 962,436 belong to the \textit{Benign} class and 107,778 to the \textit{Exfiltration} class. The target variable \texttt{label} was encoded using \textit{Label Encoding}, and the test set was aligned to the schema of the training set. For the SVM, the numerical features were additionally standardized with \textit{StandardScaler}, given the model's sensitivity to the scale of the data. The \textit{Random Forest} was configured with 100 trees, \textit{random state} equal to 50, and \texttt{class\_weight="balanced"}. \textit{XGBoost} used 300 estimators, a maximum depth of 6, a learning rate of 0.1, \texttt{subsample} of 0.8, \texttt{colsample\_bytree} of 0.8, and \textit{random state} equal to 50. The SVM was configured with a radial basis function (RBF) kernel, $C = 1.0$, \textit{gamma = scale}, and \texttt{class\_weight="balanced"}. The metrics considered were \textit{Accuracy}, \textit{Balanced Accuracy}, macro \textit{Precision}, macro \textit{Recall}, macro \textit{F1-score}, weighted \textit{F1-score}, and the confusion matrix. Table \ref{tab:comparacao_modelos} summarizes the results obtained.

\begin{table}[ht]
\centering
\caption{Comparison of the supervised experiments conducted on \textit{UFU-Do53-EXF}.}
\label{tab:comparacao_modelos}
\footnotesize
\setlength{\tabcolsep}{3pt}
\renewcommand{\arraystretch}{0.95}
\begin{tabular}{lcccccccc}
\hline
\textbf{Model} & \textbf{Acc.} & \textbf{Bal. Acc.} & \textbf{Prec.} & \textbf{Rec.} & \textbf{F1} & \textbf{W. F1} & \textbf{FP} & \textbf{FN} \\
\hline
\textit{Random Forest} & 0.999996 & 0.999981 & 0.999998 & 0.999981 & 0.999990 & 0.999996 & 0 & 4 \\
\textit{XGBoost}       & 0.999998 & 0.999991 & 0.999999 & 0.999991 & 0.999995 & 0.999998 & 0 & 2 \\
SVM                    & 0.999915 & 0.999817 & 0.999714 & 0.999817 & 0.999765 & 0.999915 & 58 & 33 \\
\hline
\end{tabular}
\end{table}


The results indicate high performance in the intra-dataset evaluation of \textit{UFU-Do53-EXF}. \textit{XGBoost} achieved the best result, with only 2 false negatives and no false positives on the test set, while \textit{Random Forest} produced 4 false negatives and, likewise, no false positives. The SVM remained competitive, but generated a larger number of residual errors. In terms of detection, the few false negatives observed suggest that the lexical, statistical, and behavioral features extracted from the DNS queries preserve relevant signals of exfiltration. However, since the experiment uses a random stratified split, the results should be interpreted as evidence of intra-dataset consistency, and not as a guarantee of generalization.

\subsection{\textit{UFU-DoH-EXF}}

The second case study evaluates \textit{UFU-DoH-EXF} in a supervised binary classification task between \textit{Benign} and \textit{Exfiltration} flows. Unlike Do53, the DoH scenario imposes lower semantic observability, since DNS queries are encapsulated in HTTPS/TLS. Thus, the evaluation uses statistical features of flow, packet direction, size, duration, volume, and temporal behavior, in line with detection approaches for encrypted DoH traffic \cite{zhan_detecting_2022}. The \textit{Random Forest}, \textit{XGBoost}, and \textit{Support Vector Machine} models were evaluated. The experiments used a stratified split of 80\% for training and 20\% for testing. Before the split, exact duplicates were removed, resulting in 1,022,257 TCP/443 flows, of which 992,111 belong to the \textit{Benign} class and 30,146 to the \textit{Exfiltration} class. The test set contains 204,452 instances, of which 198,423 are benign and 6,029 are exfiltration.

The target variable \texttt{label} was encoded using \textit{Label Encoding}. For \textit{Random Forest} and \textit{XGBoost}, the numerical features were handled with median imputation when necessary. For the Linear SVM, standardization with \textit{StandardScaler} was also applied. The \textit{Random Forest} was configured with 100 trees, \textit{random state} equal to 42, and \texttt{class\_weight="balanced"}. \textit{XGBoost} used 300 estimators, a maximum depth of 6, a learning rate of 0.1, \texttt{subsample} of 0.8, \texttt{colsample\_bytree} of 0.8, the \texttt{binary:logistic} objective, the \texttt{logloss} evaluation metric, and \texttt{scale\_pos\_weight} of approximately 32.91. The Linear SVM was configured with $C = 1.0$, \texttt{max\_iter = 10000}, and \texttt{class\_weight="balanced"}. Table \ref{tab:comparacao_modelos_doh} summarizes the results obtained.

\begin{table}[ht]
\centering
\caption{Comparison of the supervised experiments conducted on \textit{UFU-DoH-EXF}.}
\label{tab:comparacao_modelos_doh}
\footnotesize
\setlength{\tabcolsep}{3pt}
\renewcommand{\arraystretch}{0.95}
\begin{tabular}{lcccccccc}
\hline
\textbf{Model} & \textbf{Acc.} & \textbf{Bal. Acc.} & \textbf{Prec.} & \textbf{Rec.} & \textbf{F1} & \textbf{W. F1} & \textbf{FP} & \textbf{FN} \\
\hline
\textit{Random Forest} & 0.999985 & 0.999832 & 0.999912 & 0.999832 & 0.999872 & 0.999985 & 1  & 2 \\
\textit{XGBoost}       & 0.999985 & 0.999832 & 0.999912 & 0.999832 & 0.999872 & 0.999985 & 1  & 2 \\
SVM             & 0.999863 & 0.999688 & 0.997927 & 0.999688 & 0.998806 & 0.999863 & 25 & 3 \\
\hline
\end{tabular}
\end{table}


The results also indicate high performance for the DoH scenario, even without access to the semantic content of DNS queries. \textit{Random Forest} and \textit{XGBoost} produced the same confusion matrix, but did not misclassify exactly the same instances, while the Linear SVM produced more false positives. This result suggests that flow, size, direction, duration, and timing features capture observable effects of exfiltration even in encrypted traffic. Nevertheless, since this is an intra-dataset evaluation, the results should be understood as a supervised baseline for the proposed dataset. Future evaluations should consider splits by session, capture period, domain, or exfiltration tool, in order to measure generalization under scenarios closer to real-world deployment.

\section{Conclusion}

This paper presented two complementary datasets for DNS exfiltration detection: \textit{UFU-Do53-EXF} and \textit{UFU-DoH-EXF}. The main contribution of this work is methodological and experimental. Rather than treating Do53 and DoH as isolated problems, supported by heterogeneous and poorly comparable datasets, the proposal establishes a common framework for data generation, capture, and transformation, with controlled benign and malicious traffic, preservation of PCAP artifacts, reproducible conversion to network flows, explicit feature definitions, and a direct focus on the exfiltration problem.

The quantitative results show that both datasets have a relevant scale for machine-learning-based detection studies. \textit{UFU-Do53-EXF} comprises 5,351,064 samples after normalization and presents a profile suitable for supervised evaluation under class imbalance. \textit{UFU-DoH-EXF}, in turn, contains 1,048,575 TCP/443 flows, of which 1,016,438 are benign samples and 32,137 are exfiltration samples. In particular, the experimental results with \textit{UFU-DoH-EXF} show that, even without access to the semantic content of DNS queries, statistical features of flow, direction, size, duration, volume, and temporal behavior can capture observable effects of exfiltration in encrypted traffic. 


As future work, four directions stand out. The first aims to strengthen the experimental evaluation through stratified cross-validation with 5 and 10 folds, in addition to splits by session, capture period, and exfiltration tool, making it possible to distinguish the consistency of the results from their generalization capability. The second proposes incorporating and evaluating, through ablation analysis, the TLS metadata derived from the \textit{ClientHello} (SNI, ALPN, and JA3), in order to measure their influence on class separability and to verify whether detection remains based on behavioral characteristics even in scenarios with \textit{Encrypted Client Hello} \cite{rescorla_tls_2026}. The third involves evaluating the sensitivity of detection in DoH scenarios under different network and implementation conditions, including latency, jitter, resolvers, and TLS/HTTP stacks, as well as robustness tests with noise and resampling. Finally, we intend to perform validation across public and private datasets and to develop an integrated machine-learning-based architecture for DNS exfiltration detection.

\newpage

\bibliographystyle{sbc}
\bibliography{referencias}

\end{document}